\documentclass[a4paper]{article}
\usepackage{amsmath}
\usepackage{graphicx}
\begin{document}
\title{The multiplets of finite width $0^{++}$ mesons and encounters with exotics}
\author{Micha{\l} Majewski
\thanks{e-mail:m.majewski@merlin.fic.uni.lodz.pl}
\\Department of Theoretical Physics II, University of Lodz \\Pomorska 149/153, 90-236 Lodz, Poland}
\maketitle
\begin{abstract}
Complex-mass (finite-width) $0^{++}$ nonet and decuplet are
investigated by means of exotic commutator method. The hypothesis
of vanishing of the exotic commutators leads to the system of
master equations (ME). Solvability conditions of these equations
define relations between the complex masses of the nonet and
decuplet mesons which, in turn, determine relations between the
real masses (mass formulae), as well as between the masses and
widths of the mesons. Mass formulae are independent of the
particle widths. The masses of the nonet and decuplet particles
obey simple ordering rules. The nonet mixing angle and the mixing
matrix of the isoscalar states of the decuplet are completely
determined by solution of ME; they are real and do not depend on
the widths. All known scalar mesons with the mass smaller than
$2000MeV$ (excluding $\sigma(600)$) and one with the mass
$2200\div2400MeV$ belong to two multiplets: the nonet $(a_0(980),
K_0(1430), f_0(980), f_0(1710))$ and the decuplet $(a_0(1450),
K_0(1950), f_0(1370), f_0(1500), f_0(2200)/f_0(2330))$. It is
shown that the famed anomalies of the $f_0(980)$ and $a_0(980)$
widths arise from an extra "kinematical" mechanism, suppressing
decay, which is not conditioned by the flavor coupling constant.
Therefore, they do not justify rejecting the $q\bar{q}$ structure
of them. A unitary singlet state (glueball) is included into the
higher lying multiplet (decuplet) and is divided among the
$f_0(1370)$ and $f_0(1500)$ mesons. The glueball contents of these
particles are totally determined by the masses of decuplet
particles. Mass ordering rules indicate that the meson
$\sigma(600)$ does not mix with the nonet particles.

\end{abstract}

\section{Introduction}
Thirty years ago David Morgan posed a question of the
``respectability'' of scalar mesons as $q\bar{q}$ systems \cite
{Morgan}. He attempted to find an affirmative answer to this
question. Soon after that, people became sceptical about such a
possibility. Primarily, the main reason was the supposed
domination of the $f_0(980)\rightarrow{K\bar{K}}$ decay channel.
Later, after establishing that this decay is not dominating (PDG
have been announcing domination of the mode $f_0(980)\rightarrow
\pi\pi$ since 1982), the disagreement between measured
$\Gamma^{exp}$ \cite{PDG} and predicted $\Gamma^{q\bar{q}}$
\cite{G-I} total width of the decay:
\begin{equation}
    \Gamma^{exp}=40\div100\, \mathrm{MeV},      \label{1.1}
\end{equation}
\begin{equation}
    \Gamma^{q\bar{q}}=500\div1000\, \mathrm{MeV}.   \label{1.2}
\end{equation}
was recognized as a main argument against the $q\bar{q}$ structure
of the $f_0(980)$ meson. Probably, this argument was never
contested.

So a question arose as to the internal structure of the
$f_0(980)$, $a_0(980)$ and other mesons forming a scalar
multiplet. Many alternative models were created to explain this
multiplet. The most prominent ones are exotic models describing
scalar mesons totally or partly as $qq\bar{q}\bar{q}$ states.
These models differ from one another with physical interpretation
and/or construct the multiplet from different particles. It is not
the purpose of this paper to discuss these issues. In the
extensive bibliography introducing these models the interested
reader is referred to a few representative papers \cite{Jaffe},
\cite{Wein-Isg}, \cite{Achas}, \cite{Tornq}, \cite{C-T} . It
should be, however, recognized that the views of many authors
evolved remarkably during the time elapsed.  We abandon discussing
them, because we question the validity of the argument that the
disagreement between the numbers of (\ref{1.1}) and (\ref{1.2})
can be regarded as evidence against the $q\bar{q}$ nature of the
scalar mesons. Consequently, we question the exotic models of the
scalar nonet.

Although the arguments against rejection of the $q\bar{q}$ model
will be set forth later, it is worth noting here that the
disagreement between (\ref{1.1}) and (\ref{1.2}) by itself does
not certify a contradiction. The contradiction only appears if one
admits that observed width is determined entirely by flavor
coupling constant. Such a point of view is widely shared, in spite
of many examples of hadronic decays revealing additional
suppression. The reason is that there is no known mechanism which
could suppress the $f_0(980)$ decay. However, as we argue below,
such a mechanism must exist. Its indispensability is clearly seen
if the mesons are described as finite-width nonet states.

Below, we use notion of the ``flavor width''(FW) which is
distinguished from ``hadronic width'' (HW). It has been shown that
total FW which is determined by the flavor coupling constant is
reduced to an experimentally observed total HW due to some
"kinematical" suppression mechanism \cite{SR}. Thus the number
from (\ref{1.1}) is the HW, while the number from (\ref{1.2}) is
the FW of the $f_0(980)$ meson.

Another kind of exotics is being searched. According to a wide
spread opinion, there should exist a glueball at $\simeq 1.5
\mathrm{GeV}$ \cite{Ba}, \cite{Ger}, \cite{Ani}, \cite{Am-Ar},
\cite{Am-C}, \cite{C-K}, \cite{CAms}, \cite{C-Z}. This particle is
not expected to be a pure state - it should be a mixture of the
glueball and the isoscalar $q\bar{q}$ state of a nonet. That
creates a decuplet. The abundance of the scalar mesons suggests
that there exists more than one multiplet - we may expect a nonet
and a decuplet. The problem is how the particles are distributed
among them. A solution can be found if we exploit accessible
knowledge about these multiplets.

That can be achieved by means of exotic commutator method (ECM)
\cite{MT1}. Using this method a system of algebraic equations
("master equations" (ME)) was derived for the octet contents of
the isoscalar members of the zero-width meson nonet and decuplet
\cite{MT1,MT2}. Solution of the ME gives full attainable
information about these multiplets. For the nonets it clearly
distinguishes three kinds of them: Gell-Mann--Okubo (GMO),
Schwinger (S) and ideally mixed (I) ones. The differences matter
in analysis of the data. But ME gives not only the old-standing
relations, as the mass formulae and an expression for the mixing
angle of the nonet, but also something new (derived, as yet, only
in the ECM approach): the nonet and decuplet mass ordering rules,
the decuplet mass formula and the decuplet mixing matrix. The
later follows directly from the solution of the ME, without
additional assumptions which are needed in other approaches for
diagonalizing the unphysical mass operator. The mass ordering
rules help in composing the multiplets of scalar mesons and make
the description of whole collection of the scalar mesons simple
and transparent.

This method was also applied to describing a finite-width
(complex-mass) mesons. Many nonets with different $J^{PC}$ were
fitted \cite{SR}. The fits demonstrate that most of the observed
nonets are the S ones. Besides, extension of the ME to the complex
mass reveals that the widths of the S nonet mesons depend linearly
on the masses of the particles. The slope of the line is negative
for all known nonets. The linearity follows from flavour
properties of the nonet. It is broken in all observed low mass
nonets.  The mechanism of the breaking is "kinematical" - it does
not depend on the flavor coupling constant. This fact is important
for the interpretation of the suppression of the $f_0(980)$ and
$a_0(980)$ meson decays.

The part of the present paper concerning the nonet of $0^{++}$
mesons is a continuation of the previous analysis \cite{SR}. We
justify the status and confirm the properties of the scalar nonet
which were admitted to be still controversial there.

Our purpose is also to discuss the fit of the finite-width
decuplet of the $0^{++}$ mesons, but first we have to make ME
predictions for this multiplet. Therefore, we begin with recalling
the ME procedure, fixing also the notation.

\section{Exotic commutators and master equations for octet contents of the
physical isoscalar states}

The following sequence of exotic commutators is assumed to vanish
\cite{MT1}:

\begin{equation}
    \left[T_a,\frac{d^{j}T_b}{dt^j}\right]=0,\quad \left(j=1,2,3,...\right)
\label{2.1}
\end{equation}
where $T$ is $SU(3)_F$ generator, $t$ is the time and
$\left(\alpha{},\beta{} \right)$ is an exotic combination of
indices, i.e. such that the operator
$\left[T_\alpha,T_\beta\right]$ does not belong to the octet
representation. Substituting $\frac{dT}{dt}=i[H,T]$, and using the
infinite momentum approximation for one-particle hamiltonian $H =
\sqrt{m^2+p^2}$, we transform eqs.~(\ref{2.1}) into the system:
\begin{displaymath}
    [T_\alpha,[\hat{m^2},T_\beta]]=0,
\end{displaymath}
\begin{displaymath}
    [T_\alpha,[\hat{m^2},[\hat{m^2},T_\beta]]]=0,
\end{displaymath}
\begin{equation}
    [T_\alpha,[\hat{m^2},[\hat{m^2},[\hat{m^2},T_\beta]]]]=0, \label{2.2}
\end{equation}
\begin{displaymath}
........................................
\end{displaymath}
where $\hat{m^2}$ is the squared-mass operator.

For the matrix elements of the commutators (\ref{2.2}) between
one-particle states (we assume one-particle initial, final and
intermediate states) we obtain the sequence of equations involving
expressions $\langle z_8|(m^2)^j|z_8\rangle$ with different powers
$j=1,2,3,..$ ($z_8$ is the isoscalar state belonging to the
octet). Solving these equations, we obtain the sequence of
formulae for a multiplet of the light mesons. We find
\begin{equation}
    \langle{z_8}\mid{\hat{(m_c^2)}^j}\mid{z_8}\rangle = \frac{1}{3} a_c^j +
    \frac{2}{3}b_c^j \quad (j=1,2,3,...). \label{2.3}
\end{equation}
Here $\hat{m_c^2}$ is assumed to be a complex-mass squared
operator \cite{SR}:
\begin{equation}
    \hat{m_c^2}=\hat{m^2}-i\hat{m}\hat{\Gamma}. \label{2.4}
\end{equation}
This operator can be diagonalized and has orthogonal eigenvectors.
For the complex masses of the individual particles we use
following notation:
\begin{equation}
    \label{2.5}
    \begin{aligned}
    a_c&=a-i\alpha=(m_a)^2-im_a\Gamma_a,\\
    K_c&=K-i\kappa=(m_K)^2-im_K\Gamma_K,\\
    z_j&=x_j-iy_j=(m_j)^2-im_j\Gamma_j,\\
    z_8&=x_8-iy_8=(m_8)^2-im_8\Gamma_8,\\
    b_c&=b-i\beta=(m_b)^2-im_b\Gamma_b.
    \end{aligned}
\end{equation}
The symbols $a$ and $K$ mean isotriplet and isodoublet meson
respectively; $z_j$ are isoscalar mesons; the real and imaginary
parts of the subsidiary complex masses $z_8$ and $b_c$ are:
\begin{equation}
    x_8=\frac{1}{3}(4K-a), \quad y_8=\frac{1}{3}
    (4\kappa-\alpha), \label{2.6}
\end{equation}
\begin{equation}
    b=2K-a, \quad \beta=2\kappa-\alpha.   \label{2.7}
\end{equation}
Numbering of the physical isoscalar mesons $z_i$ is chosen such
that their masses obey the inequality
\begin{equation}
    x_i < x_{i+1}. \label{2.7a}
\end{equation}

The octet state $|z_8\rangle$ can be expressed by physical
isosinglet states $|z_i\rangle$. For the nonet we substitute into
(\ref{2.3}) the expression:
\begin{equation}
    |z_8\rangle=l_1|z_1\rangle + l_2|z_2\rangle,  \label{2.8}
\end{equation}
and for the decuplet:
\begin{equation}
    |z_8\rangle=l_1|z_1\rangle + l_2|z_2\rangle+l_3|z_3\rangle. \label{2.8a}
\end{equation}
The coefficients $l_i$ are complex numbers satisfying
\begin{equation}
            \Sigma |l_i|^2=1.  \label{2.9}
\end{equation}
As a result, we obtain the linear system of \it master equations
\rm (ME) determining the octet contents $|l_i|^2$  of the
isoscalar $z_i$ states:
\begin{equation}
    \Sigma |l_i|^2z_i^j=\frac{1}{3}a_c^j+\frac{2}{3}b_c^j, \quad (j=0,1,2,3,..)
    \label{2.10}
\end{equation}
where the equation for $j=0$ takes into account the condition
(\ref{2.9}). ME can be applied to analyzing the nonet and decuplet
of the real and complex-mass mesons in the broken $SU(3)_F$
symmetry. The mass formulae arise if the number of equations
exceeds the number of unknown coefficients $l_i$. They play a role
of solvability condition of the system (\ref{2.10}).

\section{Nonet of $0^{++}$ mesons - back to $q\bar{q}$}

\subsection{Three kinds of the nonets}

Solution of the ME (\ref{2.10}) for $|l_1|^2$, $|l_2|^2$ and the
mass formulae for a nonet have already been analyzed \cite{SR}. We
report on the main points of that analysis.

The system (\ref{2.10}) can be solved if the number of equations
$\geq{2}$. In such cases $|l_1|^2$, $|l_2|^2$ can be determined
from the first two of them. We find

\begin{equation}
    |l_1|^2=\frac{1}{3}\frac{(z_2-a_c)+2(z_2-b_c)}{z_2-z_1}, \label{23.1}
\end{equation}

\begin{equation}
    |l_2|^2=\frac{1}{3}\frac{(a_c-z_1)+2(b_c-z_1)}{z_2-z_1}. \label{23.2}
\end{equation}
Then the subsequent equations have to be identically satisfied.
These identities are complex mass formulae. A different number of
the mass formulae define different kinds of the nonet.

No condition of solvability and, respectively, no mass formula
exists for the system of the first two ME. This system can be
written in the form
\begin{equation}\label{23.3}
    z_1\sin^2\theta+z_2\cos^2\theta=z_8,
\end{equation}
where $\theta$ is mixing angle and $z_8$ is the Gell-Mann--Okubo
mass squared:
\begin{equation}\label{23.4}
    z_8\equiv\frac{1}{3}a_c+\frac{2}{3}b_c.
\end{equation}
The formula (\ref{23.3}) is sometimes considered as the nonet mass
formula. Such a nonet is called the GMO one. It arises for the
system of two equations (\ref{2.10}).

For the system of three equations (\ref{2.10}) we get one mass
formula:
\begin{equation}
    (a_c-z_1)(a_c-z_2)+2(b_c-z_1)(b_c-z_2)=0. \label{23.5}
\end{equation}
This is the Schwinger (S) complex-mass formula.

From the system of four equations (\ref{2.10}), besides of the
mass formula (\ref{23.5}), we obtain
\begin{equation}\label{23.6}
a_c(a_c-z_1)(a_c-z_2)+2b_c(b_c-z_1)(b_c-z_2)=0.
\end{equation}
From (\ref{23.5}) and (\ref{23.6}), choosing  the numbers of $z_i$
according to the rule (\ref{2.7a}), we get:
\begin{equation}\label{23.7}
  z_1=a_c, \quad z_2=b_c, \quad |l_1|^2=\frac{1}{3},
  \quad|l_2|^2=\frac{2}{3}.
\end{equation}
This is a complex-mass ideally mixed nonet.

The system including one more equation (\ref{2.10}) gives one more
mass formula
\begin{equation}\label{23.8}
    a_c^2(a_c-z_1)(a_c-z_2)+2b_c^2(b_c-z_1)(b_c-z_2)=0,
\end{equation}
which is satisfied by the ideally mixed nonet. It is now obvious
that also the subsequent equations of the system (\ref{2.10})
comply with ideality \cite{MT1}.

We thus find that ECM predicts just three kinds of complex-mass
nonets: GMO, S and I. Each of them defines three kinds of
connections between real quantities:

\begin{enumerate}
\item between real parts of the complex-mass squared: the mass
formulae; \item between masses and widths of the particles:
defining the flavor stitch line of the masses on the complex
plane; \item between imaginary parts of the complex-mass squared:
width sum rules.
\end{enumerate}
An important property of the mass formulae for the nonet of any
kind is their independence of the particle widths and coincidence
with respective mass formulae for the real-mass meson nonet.
Therefore, the complex-mass meson nonets may be given the names of
the real-mass ones: GMO, S and I \cite{SR}. Following the property
of the independence of the mass formulae on the particle widths
also the definitions of these nonets are independent of them; the
mesons forming different width patterns create the same nonet, if
their masses are the same. The states of the nonet of any kind
have the $q\bar{q}$ structure and this structure is stable under
anomalies of the widths. So, \it{disagreement between expected and
observed values of the $f_0(980)$ meson widths cannot be
considered as evidence against the $q\bar{q}$ nature of this
meson}. \rm

 On the
contrary, the $q\bar{q}$ structure of $f_0(980)$ meson is
confirmed, if we indicate the nonet that it belongs to.

Experimental fits show that the well known meson nonets: $1^{--}$,
$2^{++}$ and $3^{--}$ are the S ones. Also the less known:
$1^{+-}$, $1^{++}$ and $2^{-+}$ are probably the S nonets. (Only
the pseudoscalar mesons: $\pi$, $K$, $\eta$, $\eta'$ form the GMO
nonet.) That, and the identity of the $a_0(980)$ and $f_0(980)$
meson masses suggests that scalar mesons form the S nonet. We are
looking for the nonet including $a_0(980)$, $K_0(1430)$ and
$f_0(980)$ mesons. Then, the S mass formula indicates the
$f_0(1710)$ meson as the ninth member.

We recall now the main properties of the mass formula and flavor
stitch line for the S nonet. There exists one relation between the
complex masses in this case.

\subsection{Schwinger nonet mass formula for finite-width mesons}

The S mass formula can be written in the form \cite{SR}

\begin{equation}
    (a-x_1)(a-x_2)+2(b-x_1)(b-x_2)=0. \label{2.11}
\end{equation}
For the $|l|^2$ s we find

\begin{equation}
    |l_1|^2=\frac{1}{3}\frac{(x_2-a)+2(x_2-b)}{x_2-x_1}, \label{2.12}
\end{equation}

\begin{equation}
    |l_2|^2=\frac{1}{3}\frac{(a-x_1)+2(b-x_1)}{x_2-x_1}. \label{2.13}
\end{equation}
They have to satisfy condition $|l_i|^2>0$. This condition and the
mass formula (\ref{2.11}) make the particle masses comply with the
ordering rule:
 \begin{equation}
    x_1 < a < x_2 < b,  \label{2.14}
\end{equation}
or
\begin{equation}
    a < x_1 < b < x_2 .  \label{2.15}
\end{equation}
The mixing angle $\Theta^{Sch}$ is real and totally determined by
the masses:
\begin{equation}
    tan^2\Theta^{Sch}=\frac{|l_1|^2}{|l_2|^2}. \label{2.16}
\end{equation}
This formula shows that also the mixing angle does not depend of
the widths.

 The
masses fit inequality $f(1710)<b$, pointing out the mass ordering
(\ref{2.14}) for this nonet. So the mass of the $f_0(980)$ meson
must be smaller than the mass of the $a_0(980)$ one. The mass
formula is well satisfied and mixing angle $(\theta$=$33.5\pm2)^0$
is close to ideal. For $f_0(1710)$ almost pure the $s\bar{s}$
structure is predicted.

\begin{table}
\centering
\caption{The nonet of $0^{++}$ mesons. The three rows contain
masses; widths; mixing angle, mass and width ordering. Subsidiary
quantities $m_b=\sqrt{b}$ and $\Gamma_b$=$\frac{\beta}{m_b}$ are
calculated. The large value of $\Gamma_b$ reflects strong
"kinematical" suppression of $a_0$ decay. In the ordering rules:
$a$, $b$, $x_1$, $x_2$ are masses squared; $\alpha$, $\beta$,
$y_1$, $y_2$ are products of the mass and width. Masses and widths
are given in MeV. Status of the particles, notation and data are
quoted from RPP \cite{PDG}.} \label{tab:1}
\begin{tabular}{|r@{}l||c||c|c||c|c|}
\hline
&&$m_K$&$m_a$&$m_1$&$m_b$&$m_2$\\
\multicolumn{2}{|c||}{$J^{PC}$}&$\Gamma_K$&$\Gamma_a$&$\Gamma_1$&$\Gamma_b$
&$\Gamma_2$\\
\hline
\multicolumn{2}{|c||}{particles}&\multicolumn{1}{c}{$\theta^{GMO}$}&
\multicolumn{2}{c}{mass ordering}&\multicolumn{2}{c|}{width ordering}\\
\hline\hline \multicolumn{2}{|c||}{$0^{++}$}&$1412\pm 6$&$984.7\pm
1.2$&$980\pm 10$&
$1737\pm 11$&$1714\pm 5$\\
\cline{1-2}
$\bullet$&$a_0(980)$&&&&&\\
$\bullet$&$K_0(1430)$&$294\pm 23$&$50\div 100$&$40\div 100$&
$380\div 490$&$140\pm 10$\\
$\bullet$&$f_0(980)$&&&&&\\
\cline{3-7} $\bullet$&$f_0(1710)$&\multicolumn{1}{c}{$(33.5\pm
2.0)^\circ$}& \multicolumn{2}{c}{$x_1<a<x_2<b$}&
\multicolumn{2}{c|}{$y_1>\alpha>y_2>\beta  $}\\
\hline
\end{tabular}
\end{table}




%



Thus the scalar nonet looks quite ordinary. One may wonder,
however, why it is so distinct from the other ones. This question
can be explained to some extent by inspection of its flavor stitch
line.

\subsection{The flavour stitch line}
The total widths of all physical mesons  belonging to the S nonet,
as well as the subsidiary states $z_8$ and $b_c$, satisfy the
equation:
\begin{equation}
    \frac{\Gamma_k-\Gamma_l}{m_k-m_l}   =k_s   \label{2.17}
\end{equation}
where $k$ and $l$ ($l\neq k$) run over $a$, $K$, $x_1$, $x_2$,
$x_8$, and $b$. Equation (\ref{2.17}) represents a straight line
in the $(m,\Gamma)$ plane; $k_s$ is its slope. Rectilinearity is
an effect of requiring the ME (\ref{2.10}) to hold for complex
masses, i.e. that the relations between widths, likewise between
the masses, are completely determined by flavor interaction. The
points $(m,\Gamma)$, representing the complex masses of the
individual particles, form a sequence of stitches along the
straight line. We call this the \it{flavor stitch line} \rm(FSL).

Nonets of physical particles obey the equation of the flavor
stitch line when sufficiently much decay channels are opened, so
that the $\Gamma$s are not sensitive to suppression of single mode
of decay. The data suggest that this happens as all masses of the
nonet are bigger than $\approx 1.5 \mathrm{GeV}$. We refer to the
width determined from FSL as the \it{flavor width}  \rm(FW) of a
meson. In the S nonet of the less massive particles, some of them
may have the widths reduced by additional "kinematical"
mechanism(s) suppressing decay. The outcome is the \it{hadronic
width} \rm(HW) - the quantity which is observed in experiment. The
difference between FW and HW is a measure of an extra suppression
of the decay. Disagreement between (\ref{2.17}) and the data is
best seen on the mass-width diagram. In this diagram FSL is a
straight line. The point $(m,\Gamma)$ of a physical state lying
below FSL exhibits the particle having reduced width due to a
"kinematical" mechanism. Probably in most cases it is really
kinematical (conservation laws, selection rules, phase space
etc.), but there are also possible other (known or unknown)
mechanisms of suppression; among them, also the dynamical ones. In
the same way the point $(m,\Gamma)$ of the physical particle lying
over FSL would exhibit the enhanced decay. Obviously, definition
of the S nonet is invariant upon ``kinematical `` suppression.

The slope $k_s$ of the FSL is not predicted by the model. It can
be determined only with the help of experimental data. From the
nonets where data are sufficient, we find
\begin{equation}
    k_s=-0.5\pm 0.1   \label{2.18}
\end{equation}
The value of $k_s$ is firmly determined when all points
$(m,\Gamma)$ of physical particles lie on a straight line. Also we
can approximately (not so definite) determine it, using data on
two particles, if we are convinced that their decays are not
suppressed. One of the reasons for such conviction is $m>1.5
\mathrm{GeV}$.

\begin{figure}
  \includegraphics[width=\textwidth,angle=-90]{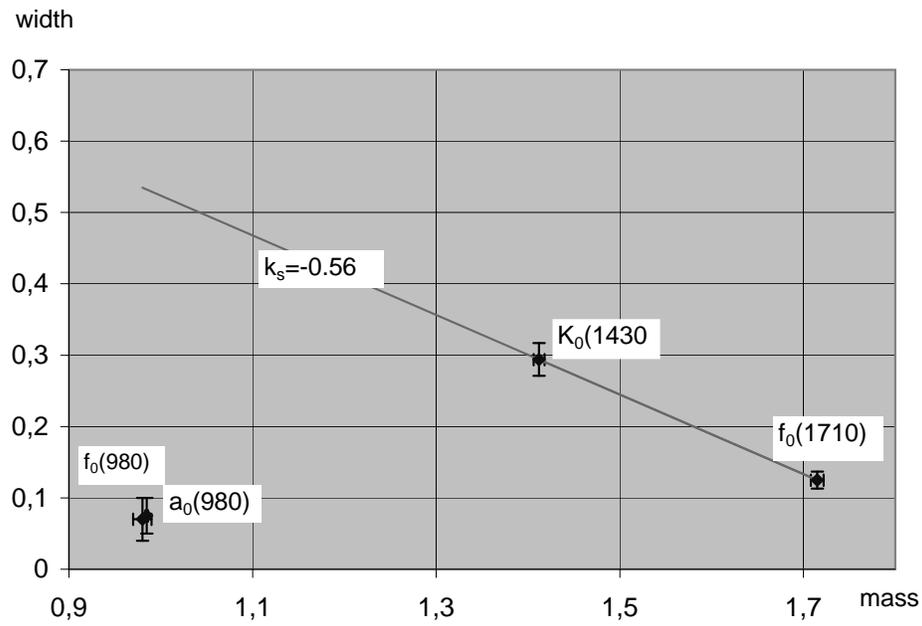}\\
  \caption{Mass-width diagram of the $0^{++}$ nonet. On the axes $m$ and $\Gamma$ are given
  in GeV. The approximate flavor stitch line is drawn according to the  coordinates of the $K_0(1430)$
  and $f_0(1710)$ mesons. The large deficit of the $f_0(980)$ and $a_0(980)$ widths demonstrates
  strong "kinematical" suppression of their decays. Equal rates of the suppression
  emphasize its flavor independence.}\label{fig:0++}
\end{figure}

The mass-width diagram of the $0^{++}$ nonet is shown on the
Fig.~\ref{fig:0++}.
An approximate FSL is determined by $(m,\Gamma)$ coordinates of
the  $K_0(1430)$ and $f_0(1710)$ mesons. Its slope, $k_s=-0.56$,
is typical for the S nonets (\ref{2.18}). The FW of $f_0(980)$
meson, $\Gamma\approx{535}\mathrm{MeV}$, is consistent with the
once predicted width $\Gamma^{q\bar{q}}$ of $f_0(980)$ meson
(\ref{1.2}). According to the Fig. 1, the FW of the $a_0(980)$
meson is the same. The observed HW are $40 \div100$ $\mathrm{MeV}$
for $f_0(980)$ and $50 \div100$ $\mathrm{MeV}$  for $a_0(980)$; so
their HW and suppression rates are also the same. Equality of the
suppression rates of the isovector and isoscalar meson decays is
quite exceptional in the low lying nonets. In other nonets (except
of the "kinematically" unsuppressed $3^{--}$) these rates are
different. This suggests that for both, the $a_0(980)$ and the
$f_0(980)$ mesons works the same suppression mechanism and so it
is isospin independent. Another feature of this mechanism is that
it does not change the masses, but this only confirms the
"kinematical" nature of the suppression.

So what can be the physical nature of the suppression? Our
analysis does not answer this question. Perhaps we should turn to
those effects which, according to current opinion, can modify
properties of the scalar mesons like confinement, vacuum effects,
violation of the chiral symmetry...

\section{Decuplet of mesons. Glueball mixing}
\subsection{The masses, widths and flavour stitch line}

The decuplet of the meson states is a reducible representation of
$SU(3)_F$ symmetry:
\begin{equation}
    \bf{10 =8\oplus1\oplus1.}\rm     \label{3.1}
\end{equation}
It arises by joining an additional singlet with a nonet. Below,
the singlet is considered as a glueball and the nonet as
$q\bar{q}$ system, but it is not necessary to specify them. We
call the latter a ``basic nonet'' of the decuplet.

ECM gives the unique possibility of a simple and transparent
description of the multiplets of complex-mass mesons. The
description of a decuplet is, in essence, identical with the
description of the nonet and is based on the same ME (c.f.
(\ref{2.3})), but with the octet state $|z_8\rangle$ given by
(\ref{2.8a}). We thus have to solve the system of linear equations
\begin{equation}
    |l_1|^2z_1^j+|l_2|^2z_2^j+|l_3|^2z_3^j=\frac{1}{3}a_c^j+
    \frac{2}{3}b_c^j \quad  /j=0,1,2,.../  \label{3.2}
\end{equation}
with respect to $|l_i|^2$s (c.f. (\ref{2.10})). The solution can
exist if the number of equations is three or more. We postulate
four equations, i.e. the vanishing of three exotic commutators. A
nonet satisfying this system of exotic commutators is ideal; in
the decuplet, the ideal structure of the basic nonet states is
violated due to mixing with the glueball. We say that the basic
nonet of the decuplet is ideal.

For the system of four equations (\ref{3.2}) we have one
solvability condition relating the complex masses of the decuplet.
It will be seen further that this relation, along with the
requirement of positivity of $|l_i|^2$s, leads to the mass
ordering rule, as another necessary condition of solvability.  The
ordering rule and the mass formula help much in completing the
decuplet, making the procedure simple and transparent.

The solution of the (\ref{3.2}) is
\begin{eqnarray}
    |l_1|^2&=&\frac{1}{3}\frac{(z_2-a_c)(z_3-a_c)+2(z_2-b_c)(z_3-b_c)}
    {(z_1-z_2)(z_1-z_3)}, \quad  \label{3.3a} \\
    |l_2|^2&=&\frac{1}{3}\frac{(z_1-a_c)(z_3-a_c)+2(z_1-b_c)(z_3-b_c)}
    {(z_2-z_1)(z_2-z_3)},  \quad   \label{3.3b}\\
    |l_3|^2&=&\frac{1}{3}\frac{(z_1-a_c)(z_2-a_c)+2(z_1-b_c)(z_2-b_c)}
    {(z_3-z_1)(z_3-z_2)}, \label{3.3c}
\end{eqnarray}
provided the complex masses of the particles satisfy the equation:
\begin{equation}
    M\stackrel{def}{=}(z_1-a_c)(z_2-a_c)(z_3-a_c)+
    2(z_1-b_c)(z_2-b_c)(z_3-b_c)=0. \label{3.4}
\end{equation}
The $|l_i|^2$s must be real numbers. It can easily be seen that
all ${Im|l_i|^2} = 0$, if the equations (c.f. (\ref{2.17}))
\begin{equation}
    \frac{y_i-y_j}{x_i-x_j}=\frac{y_i-\alpha}{x_i-a}=
    \frac{y_i-\beta}{x_i-b}=\frac{\alpha-\beta}{a-b}=k_s  \label{3.5}
\end{equation}
are satisfied for all $i$,$j$ ($j\neq{i}$) running over $z_1$,
$z_2$, $z_3$ and $z_8$.

These equations define the FSL of the complex-mass meson decuplet.
The points ($m^2,m\Gamma$) of all mesons belonging to the
decuplet, as well as of the subsidiary states $z_8$ and $b_c$, lie
on the straight line with slope $k_s$ in the ($m^2,m\Gamma$)
plane. Obviously, the points ($m,\Gamma$) corresponding  to these
mesons lie in the plane ($m,\Gamma$) on the straight line with the
same slope.

The last equation (\ref{3.5}) shows that $k_s$ can be defined by
the parameters of the $a$ and $K$ mesons: the slope of the
decuplet FSL is identical with the slope of the basic nonet one;
joining the additional singlet does not change the slope of FSL,
nor the stitch line. Also the other properties of the decuplet and
the basic nonet are the same:
\begin{enumerate}
\item linearity follows from flavor symmetry and departures from
it are result of "kinematical" suppression (enhancement) of the
decay, and \item the slope $k_s$ is not predicted and can only be
determined by data.
\end{enumerate}
Using (\ref{3.5}) we transform (\ref{3.3a}) - (\ref{3.3c}) into
\begin{eqnarray}
    |l_1|^2&=&\frac{1}{3}\frac{(x_2-a)(x_3-a)+2(x_2-b)(x_3-b)}
    {(x_1-x_2)(x_1-x_3)}, \quad  \label{3.6a} \\
    |l_2|^2&=&\frac{1}{3}\frac{(x_1-a)(x_3-a)+2(x_1-b)(x_3-b)}
    {(x_2-x_1)(x_2-x_3)},  \quad   \label{3.6b}\\
    |l_3|^2&=&\frac{1}{3}\frac{(x_1-a)(x_2-a)+2(x_1-b)(x_2-b)}
    {(x_3-x_1)(x_3-x_2)}. \label{3.6c}
\end{eqnarray}
They coincide with the $|l_i|^2$s for zero-with meson decuplet
\cite{MT2}-\cite{In}.

Let us define now the two real functions:
\begin{equation}
    M_R\stackrel{def}{=}(x_1-a)(x_2-a)(x_3-a)+ 2(x_1-b)(x_2-b)(x_3-b),
    \label{3.7}
\end{equation}
 \begin{equation}
    M_I\stackrel{def}{=}(y_1-\alpha)(y_2-\alpha)(y_3-\alpha)+
    2(y_1-\beta)(y_2-\beta)(y_3-\beta). \label{3.8}
\end{equation}
From (\ref{3.5}) it follows that
\begin{equation}
    M_I=k_s^3M_R.  \label{3.9}
\end{equation}

Due to eq. (\ref{3.5}), the real and imaginary parts of the
solvability condition (\ref{3.4}) can be written in the form:
\begin{eqnarray}
    Re M&=&M_R(1-3k_s^2)=0,\\  \label{3.10}
    ImM&=&M_I(1-\frac{3}{k_s^2})=0.  \label{3.11}
\end{eqnarray}
If $k_s^2\neq\frac{1}{3}$, then $M_R=0$. The same is true also for
$k_s^2=\frac{1}{3}$, as can be seen from (\ref{3.11}) and
(\ref{3.9}). Returning to the definition of $M_R$ (\ref{3.7}), we
find explicit form of the decuplet mass formula for finite-width
mesons:
\begin{equation}
    (x_1-a)(x_2-a)(x_3-a)+2(x_1-b)(x_2-b)(x_3-b)=0. \label{3.12}
\end{equation}
It does not depend on the particle widths and is identical with
the mass formula for zero-width mesons \cite {KM}.

The equation $M_I=0$ determines a sum rule for the decuplet
particle widths:
\begin{equation}
    (y_1-\alpha)(y_2-\alpha)(y_3-\alpha)+2(y_1-\beta)(y_2-\beta)(y_3-\beta)=0.
    \label{3.13}
\end{equation}
This equation is satisfied only if all points $(m,\Gamma)$ lie on
the FSL.

The right-hand parts of (\ref{3.6a})-(\ref{3.6c})  must be
positive. This cannot be fulfilled for arbitrary masses.
Therefore, the requirement of positivity restricts masses. These
restrictions along with the mass formula lead to the rule of the
mass ordering as the necessary condition of solvability of the
system (\ref{3.2}) \cite{MT2}-\cite{In}:
\begin{equation}
    x_1<a<x_2<b<x_3.  \label{3.14}
\end{equation}
For the imaginary parts of the complex masses we have:
\begin{equation}
    y_1<\alpha<y_2<\beta<y_3.  \label{3.15}
\end{equation}

The basic nonet of the decuplet could also be chosen as an S one.
The S nonet follows from the assumption that two exotic
commutators (\ref{2.1}) vanish. Then we would have three ME, just
enough for determining $|l_i|^2$s. As we know from fits of the
meson nonets \cite {SR}, the S nonet is not very different from
the I one and we hope that the properties of the decuplets built
on them are not very different. However, choosing the S basic
nonet we would be left without the mass formula and the ordering
rule. Therefore, we do not discuss this scheme.

To finish this section, let us note that there are no more types
of decuplet in the ME approach. For five ME (\ref{3.2}) there
arise two complex-mass formulae; beside of (\ref{3.12}), we obtain
\begin{equation}\label{3.15a}
    a_c(z_1-a_c)(z_2-a_c)(z_3-a_c)+2b_c(z_1-b_c)(z_2-b_c)(z_3-b_c)=0.
\end{equation}
These formulae define the ideal nonet and disconnected unitary
singlet with arbitrary mass. This result does not change if we
join the next ME (\ref{3.2}).

\subsection{Completing the decuplet}
Completing the decuplet is quite easy due to mass ordering rule
and simplicity of the mass formula.

The decuplet should include the isoscalar components  not
belonging to the nonet: $f_0(1370)(\equiv z_1)$ and
$f_0(1500)(\equiv z_2)$. The later one is considered by many
authors \cite{Ani} - \cite{C-Z} as the most likely glueball
candidate. In the same mass region we find isotriplet meson
$a_0(1450)$. If these three mesons belong to the same multiplet,
irrespectively to the nonet or to the decuplet, then, according to
the mass ordering rule, their masses should obey inequalities:
\begin{equation}
    x_1<a<x_2.     \label{3.16}
\end{equation}

They could belong to the nonet, if there existed a $K_0$ meson
with such mass that
\begin{equation}
    b=2K_0-a_0(1450)  \label{3.17}
\end{equation}
(c.f. (\ref{2.7})) not much exceeds $x_2$. Then the meson
$f_0(1500)$ would have structure close to $s\bar{s}$. Such a $K_0$
meson is not observed. On the other hand, nobody expects
$f_0(1500)$ to have the $s\bar{s}$ structure. Therefore, the
needed meson $K_0$ should have higher mass and the multiplet
should be a decuplet.

The only $K_0$ candidate which may play this role is $K_0(1950)$,
still ``needing confirmation''. If it is accepted, then only one
isoscalar meson $f_0$ is lacking to complete the decuplet. At
present, several signals are announced \cite{PDG}:
\begin{eqnarray}
f_0(2020),\; f_0(2060),\; f_0(2100),\; f_0(2200),\; f_0(2330).
\label{3.17a}
\end{eqnarray}
The tenth candidate should be pointed out by the value of the mass
calculated from the mass formula (\ref{3.12}). However, on account
of large error of most of the input masses, we first perform
qualitative discussion of the formula.

An exceptionally large difference between the masses of the
$a_0(1450)$ and $K_0(1950)$ mesons, $m_K-m_a \simeq 500\
\mathrm{MeV}$, enables us to estimate very precisely the
difference $x_3-b$. (Notice, by the way, that such a large
difference between appropriate masses is observed also in the
nonet $0^{++}$). According to the mass ordering rule, this
difference must be positive. For such a large mass of $K_0$ the
difference $b-a$ is also large ($b -a \simeq 5.5\,
\mathrm{GeV}^2$) and we have:
\begin{equation}
    x_3-a\approx b-a,\;\;\;b-x_1\approx b-x_2\approx b-a. \label{3.18}
\end{equation}
Then, from (\ref{3.12}) we find:
\begin{equation}
    x_3=b+\frac{(x_2-a)(a-x_1)}{2(b-a)}.   \label{3.19}
\end{equation}
Following the opinion of \cite{Ani} - \cite{C-Z} that the glueball
dominates the structure of $z_2~(\equiv~f_0(1500))$ meson, we
admit that the light quarks dominate the structure of $z_1( \equiv
f_0(1370))$ meson; the mass of $z_1$ should be closer to the mass
of $a_0$ than the mass of $z_2$ and the masses of $a_0$, $z_1$,
$z_2$ mesons would satisfy inequality
\begin{equation}
    a-x_1<x_2-a.  \label{3.20}
\end{equation}
So we find
\begin{equation}
    b<x_3<b+\frac{(x_2-a)^2}{2(b-a)},  \label{3.21}
\end{equation}
 or
\begin{equation}\label{3.21a}
    x_3-b<1 MeV^2.
\end{equation}
The poorly known masses of $K_0(1950)$ and $f_0$ mesons appear to
be strongly correlated.

The particles creating decuplet are mentioned in the Tab. 2. The
last column shows the mass of the $b$-state $m_b$. The value of
this mass points out $f_0(2330)$ as the best candidate for the
heaviest isoscalar of the decuplet.

The mass of the $b$-state shown in Tab. 2 has a large uncertainty
due to the error of the $a_0(1450)$ mass and, especially, to the
uncertainty of the $K_0(1950)$ mass. Therefore, it may be
interesting to admit another particle - the $f_0(2200)$ - as the
third isoscalar component and construct an adequate decuplet. We
can do this, keeping $f_0(1500)$ the mostly glueball state with
unchanged mass (the best known mass in the decuplet) and allowing
to change the other masses as to obey the mass formula
(\ref{3.12}).

Tab. 3 shows these two possible decuplets. Besides the masses of
the particles, are also shown the octet contents of the isoscalar
mesons. A common feature of these two solutions of the ME is the
small value of $|l_2|^2$. This follows from the assumption that
$f_0(1500)$ meson is mostly the glueball state. The mixing
matrices belonging to these solutions are presented in the next
section and denoted by $V_1$ and $V_2$.
\begin{table}
\centering
\caption{The decuplet of scalar mesons. The decuplet is formed out
of the mesons satisfying decuplet mass formula (\ref{3.12}). Three
of them are well known. The remaining two ($K_0$ and $f_0(2330)$)
are not firmly established. Their masses are strongly correlated
which supports them mutually as candidates to the decuplet. The
predicted ordering rules for masses and widths are given in the
last row. The width ordering rule cannot be verified with present
data. For notations, see the caption of the Tab.~\ref{tab:1}.}
\label{tab:2}
\begin{tabular}{|r@{}l||c||c|c||c|c|c|} \hline
&&$m_K$&$m_a$&$m_1$&$m_2$&$m_3$&$m_b$\\
\multicolumn{2}{|c||}{$0^{++}$}&$\Gamma_K$&$\Gamma_a$&$\Gamma_1$&$\Gamma_2$
&$\Gamma_3$&$\Gamma_b$\\
\hline \multicolumn{2}{|c||}{particles}&\multicolumn{1}{c}{}&
\multicolumn{2}{c}{mass ordering}&\multicolumn{3}{c|}{width ordering}\\
\hline\hline
$\bullet$&$a_0(1450)$&$1945\pm 30$&$1474\pm 19
$&$1200\div 1500$&
$1507\pm 5$&$2330$&$2325\pm92$\\
&$K_0(1950)$&&&&&&\\
$\bullet$&$f_0(1370)$&$201\pm 113$&$265\pm13$&$200\div 500$&
$109\pm7$&$220$&\\
$\bullet$&$f_0(1507)$&&&&&&\\
\cline{3-8} &$f_0(2330)$& \multicolumn{3}{c}{$x_1<a<x_2<b<x_3$}&
\multicolumn{3}{c|}{$y_1>\alpha>y_2>\beta>y_3 $}\\
\hline
\end{tabular}
\end{table}

\begin{table}
\centering
\caption{Two possible solutions of the ME (\ref{3.12}) for
decuplet adequate to choice of $f_0(2330)$ and $f_0(2200)$ meson
as the heaviest isoscalar state. In both cases the solution is
chosen such that $f_0(1500)$ is mostly glueball. One can see that
the small content of the octet state is the signature of glueball.
Masses are given in MeV. Notations and data are quoted from RPP
\cite{PDG}.} \label{tab:3}
\begin{tabular}{|r@{}l||c|c|c|c|c|}
\hline
\multicolumn{2}{|c||}{Solution}&$m_K$&$m_a$&$m_1$&$m_2$&$m_3$\\
\multicolumn{2}{|c||}{number}&&&$|l_1|^2$&$|l_2|^2$
&$|l_3|^2$\\\hline

\hline\multicolumn{2}{|c||}{1}&$1945.0$&$1474.0$&$1465.0$&
$1505.98$&$2322.45$\\
&&&&$0.25668$&$0.07691$&$0.66642$\\

\hline \multicolumn{2}{|c||}{2}&$1870.0$&$1460.0$&$1443.0$&
$1507.66$&$2205.34$\\
&&&&$0.23841$&$0.09590$&$0.66569$\\

\hline
\end{tabular}
\end{table}

\subsection{Mixing matrix}
Joining the glueball state with the quark nonet states rises the
problem of constructing the mixing matrix for three isoscalar
states.  This problem was so far formulated only for zero-width
mesons. In this case, the mixing matrix is obtained by
diagonalizing some postulated unphysical mass operator. The most
general form of this operator is simplified by additional
assumption(s) which reduce the number of independent parameters,
to facilitate the diagonalization \cite{Freund}. That makes the
result model dependent. It would be still more difficult to obtain
any result for the finite-width meson decuplet in this way.

ECM provides another procedure of constructing the mixing matrix.
It is based on the solution of the ME (\ref{2.10}) given by
$|l_1|^2$, $|l_2|^2$, $|l_3|^2$ (\ref{3.6a}) - (\ref{3.6c}),
defining the octet contents of the isoscalar mesons \cite{MT2,
In}. There is no need for introducing the mass operator nor making
assumptions about the mixing mechanism, except the natural
conjecture about flavor independence of the glueball. The octet
contents are expressed by physical masses and nothing else. The
method enables one to construct equally easy the mixing matrix
both for zero-width particles and finite-width ones. We will
calculate it in the latter case.

Let us introduce mixing matrix U transforming isoscalar states of
exact symmetry $SU(3)_F$ into the physical ones:
\begin{equation}\label{3.22}
 \begin{bmatrix}
   |z_1\rangle\\
   |z_2\rangle\\
   |z_3\rangle\\
 \end{bmatrix}   =U\begin{bmatrix}
   |z_8\rangle\\
   |z_0\rangle\\
   | G \rangle\\
 \end{bmatrix}
\end{equation}
where $z_0$ is $q\bar{q}$ singlet and $G$ is a glueball. For
complex-mass particles the matrix is, in general, unitary:

\begin{equation}\label{3.23}
    U=\begin{bmatrix}
     c_1 & -s_1c_2 & s_1s_2 \\
      s_1c_3 & c_1c_2c_3-s_2s_3e^{i\delta} & -c_1s_2c_3-c_2s_3e^{i\delta}\\
      s_1s_3 & c_1c_2s_3+s_2c_3e^{i\delta} & -c_1s_2s_3+c_2c_3e^{i\delta}\\
    \end{bmatrix}.
\end{equation}
Here $c_j$ = cos $\vartheta_j$; $s_j$ = sin $\vartheta_j$;
($j=1,2,3$); $\vartheta_j$ are Euler angles:
$0\leq\vartheta_1<\pi$; $0\leq\vartheta_2$, $\vartheta_3<2\pi$;
$\delta$ is an arbitrary phase. The elements of the first column
are $l_1$, $l_2$, $l_3$ i.e., the coefficients which were
introduced in (\ref{2.8a}). The squared absolute values of these
coefficients are solution (\ref{3.6a})-(\ref{3.6c}) of the system
(\ref{3.2}). Therefore, we have:
\begin{equation}
    c_1=\pm\sqrt{|l_1|^2}; \;\; s_1c_3=\pm\sqrt{|l_2|^2}; \;\;
    s_1s_3=\pm\sqrt{|l_3|^2}.  \label{3.24}
\end{equation}
Thus the angles $\vartheta_1$ and $\vartheta_3$ are determined by
the masses up to the signs of $c_1, c_3, s_3$.

To compare the predictions with data, the mixing matrix is usually
expressed in the basis of the ideal nonet states
\begin{equation}\label{3.29}
    |N\rangle=\frac{1}{\sqrt{2}}|(u\bar{u}+d\bar{d})\rangle, \;\;
    |S\rangle=|s\bar{s}\rangle.
\end{equation}
The physical isoscalar states are:
\begin{equation}\label{3.30}
 \begin{bmatrix}
   |z_1\rangle \\
   |z_2\rangle \\
   |z_3\rangle \\
 \end{bmatrix}=V\begin{bmatrix}
     |N\rangle \\
     |S\rangle \\
     |G\rangle \\
 \end{bmatrix},
\end{equation}
where
\begin{equation}\label{3.31}
    V=UQ,
\end{equation}
and the matrix Q
\begin{equation}\label{3.32}
    Q=\begin{bmatrix}
      \frac{1}{\sqrt{3}} & -\sqrt{\frac{2}{3}} & 0 \\
      \sqrt{\frac{2}{3}} & \frac{1}{\sqrt{3}}& 0 \\
      0 & 0 & 1 \\
    \end{bmatrix},
\end{equation}
transforms the two bases
\begin{equation}\label{3.33}
    \begin{bmatrix}
      |z_8\rangle \\
      |z_0\rangle \\
      |G\rangle
      \\
    \end{bmatrix}=Q\begin{bmatrix}
      |N\rangle \\
      |S\rangle \\
      |G\rangle \\
    \end{bmatrix}.
\end{equation}
The explicit form of the matrix $V$ is:


\begin{equation}\label{3.34}
   V=\begin{bmatrix}
      \frac{1}{\sqrt{3}} c_1-\sqrt{\frac{2}{3}} s_1c_2 & -\sqrt{\frac{2}{3}}c_1- \frac{1}{\sqrt{3}}s_1c_2& s_1s_2 \\
     \frac{1}{\sqrt{3}}s_1c_3+\sqrt{\frac{2}{3}}(c_1c_2c_3-s_2s_3e^{i\delta}) & -\sqrt{\frac{2}{3}}s_1c_3+ \frac{1}{\sqrt{3}}(c_1c_2c_3-s_2s_3e^{i\delta})& -c_1s_2c_3-c_2s_3e^{i\delta}\\
      \frac{1}{\sqrt{3}}s_1s_3+\sqrt{\frac{2}{3}}(c_1c_2s_3+s_2c_3e^{i\delta}) & -\sqrt{\frac{2}{3}}s_1s_3+\frac{1}{\sqrt{3}}(c_1c_2s_3+s_2c_3e^{i\delta})& -c_1s_2s_3+c_2c_3e^{i\delta}\\
   \end{bmatrix}
\end{equation}
The angle $\vartheta_2$ can also be determined, if flavour
independence of the glueball is imposed:
\begin {equation}
    \langle u\bar{u}|m^2|G\rangle =\langle d\bar{d}|m^2|G\rangle=
    \langle s\bar{s}|m^2|G\rangle.  \label{3.25}
\end{equation}
 For the state $|z_8\rangle$ this reads:
\begin{equation}
    \langle z_8|m^2|G\rangle =0.  \label{3.26}
\end{equation}
Substituting here $|z_8\rangle$ and  $|G\rangle$, expressed from
(\ref{3.22}) and (\ref{3.23}), we find the following equation for
$\vartheta_2$:
\begin{equation}\label{3.27}
    \tan\vartheta_2=e^{-i\delta}\frac{c_3s_3}{c_1}\frac{z_3-z_2}{(z_3-z_1)-(z_3-z_2)c_3^2}.
\end{equation}
Separation of the real and imaginary parts of $z_i$ leads, after
obvious modifications, to the same equation, with $z_i$ replaced
by $x_i$. In this equation, $e^{-i\delta}$ is the only complex
number. Therefore, $\delta=0$ and $\vartheta_2$ is completely
determined by the masses:
\begin{equation}\label{3.28}
    \tan\vartheta_2=
    \frac{c_3s_3}{c_1}\frac{x_3-x_2}{(x_3-x_1)-(x_3-x_2)c_3^2}.
\end{equation}

V is now orthogonal matrix. It is independent of the particle
widths, has no free parameters and is identical to the mixing
matrix of the isoscalar zero-width particle states \cite{MT2, In}.

Thus only signs of the trigonometric functions $c_j, s_j$ remain
to determine. We can choose them in the following way. Three
elements of the mixing matrix: $V_{11}, V_{13}, V_{32}$ may be
chosen positive. Then,

    $s_1>0$, since $0\leq\vartheta_1<\pi$;

    $c_1>0$, $c_2<0$, if we expect $|z_1\rangle\approx|N\rangle$
    with $V_{11}>0$;

    $s_2>0$, as $V_{13}>0$;

    $s_3<0$, $c_3>0$, if we expect $|z_2\rangle \approx |S\rangle$
    with $V_{32}>0$.\\
This choice of signs is consistent with (\ref{3.28}).

The mixing matrices for the solutions mentioned in the Tab. 3 read
\begin{equation}\label{3.34}
    V_1=\begin{bmatrix}
      0.88472 & 0.00510 & 0.46609 \\
      0.46612 & -0.01005 & -0.88466 \\
      0.00018 & 0.99994 & -0.01127 \\
    \end{bmatrix},
\end{equation}

\begin{equation}\label{3.35}
    V_2=\begin{bmatrix}
      0.86100 & 0.01081 & 0.50849 \\
      0.50861 & -0.01964 & -0.86077 \\
      0.00069 & 0.99975 & -0.02241 \\
    \end{bmatrix}.
\end{equation}

\subsection{Properties of the solutions}
Solution of the ME for a decuplet consists of a mass formula
connecting five masses and three expressions for $|l_i|^2$ which
determine octet contents of the three isoscalar components. The
later are used for constructing the mixing matrix. This approach
has been already applied for investigating the decuplets of the
zero-width mesons \cite{MT2}, \cite{KM}, \cite{In}, \cite{isn't},
\cite{Aj-Wa-Ka}, \cite{Dubna}, \cite{Prot}.
It remains unchanged for the finite-widths mesons.

Specifically for the $0^{++}$ decuplet, the properties of the
solution are dominated by the large difference between the masses
of the $a_0(1450)$ and $K_0(1950)$ mesons. This implies
$x_3\simeq{b}$ and enables us to make motivated choice of $f_0$
from among (\ref{3.17a}). The Tab. 3 and the matrices $V_1$ and
$V_2$ display two solutions of ME corresponding to different
$f_0$. The solutions confirm connection between the range of
indefiniteness of the mass of $K_0(1950)$ and the mass range
$2200\div2400$ of $f_0$. The predicted properties of the solutions
are similar, because the input masses of $f_0$ differ from one
another much less than the masses of $a_0(1450)$ and $K_0(1950)$
mesons. The Tab. 2 and the matrices $V_1$ and $V_2$ also show that
it would be difficult to make choice between the signals
(\ref{3.17a}), based only on the properties of the mixing matrix.

Another consequence of the large difference between the masses of
$a_0(1450)$ and $K_0(1950)$ is that the third $f_0$ is a pure
$s\bar{s}$ state. Therefore, the G state is included only into the
$f_0(1370)$ and $f_0(1500)$ states. Its distribution is determined
by the relations between the masses of the three mesons:
$f_0(1370)$, $a_0(1450)$ and $f_0(1500)$. Precise knowledge of
these masses is sufficient for complete determination of the
mixing matrix. As the present data are not accurate enough, we are
guided in constructing the matrix $V_1$ and $V_2$ by qualitative
suggestion that G is contained mainly in the state of $f_0(1500)$.
However, with the present data it is also possible, with the
suitable choice of the masses, that G is contained mainly in the
$f_0(1370)$-state.

\section{Summary and discussion}
In this paper we discuss only the flavor properties of the scalar
mesons imposed by broken $SU(3)_F$ symmetry. We extract as many
predictions of this symmetry as possible and verify their
consistency with the data. We neither consider the quark dynamics,
nor structure of the particles; in particular, we do not discuss
the problem of the structure of the higher lying multiplet
(decuplet) (are they excited $q\bar{q}$ states, hybrid $q\bar{q}g$
states or something else \cite{G-I}, \cite{Gold}?). Our approach
does not require information about the structure of the components
of the multiplet, but we hope that it may help in determining
them.

The predictions of the flavor symmetry can be obtained by means of
the exotic commutator method (ECM) of breaking the unitary
symmetry. The requirement of disappearing of the matrix elements
of these commutators between one-particle octet states gives the
system of master equations (ME) which determine the octet contents
of the isoscalar physical states. The ME include all information
attainable for the multiplets of $0^{++}$ mesons. By solving them,
we obtain not only all relations for these multiplets that are
already known, but also the new ones. In particular, we get the
relations for the nonet and decuplet of the finite-width mesons
which were unknown before. The only parameters of the ME are
physical masses and widths of the mesons. Therefore, the
predictions do not depend on free parameters, additional
assumptions or unphysical quantities. ECM provides the unique
possibility for investigating the implications of the flavor
symmetry in its pure and separated form.

ECM distinguishes three types of zero-width nonets:
Gell-Mann--Okubo (GMO), Schwinger (S) and ideal (I). They are
defined as satisfying the corresponding mass formulae. For the
finite-width nonets the mass formulae are also independent of the
widths of the particles and are identical with those for the
zero-width mesons. Therefore, we may keep the same definitions and
names for them. Data show that all (with one exception) known
nonets are of the type S. We consider this observation as an
experimental fact and use it for choosing the candidate to the
scalar nonet.

The mesons $a_0(980)$, $f_0(980)$ and $K_0(1430)$ are natural
candidates to the nonet. The S mass formula singles out the
$f_0(1710)$ as the ninth member of the nonet. This does not
contradict the known properties of this particle, because  it is
recognized as an $s\bar{s}$ state \cite{C-Z}. So these particles
form usual S nonet. Such an assignment cannot be affected by the
width anomalies of the $f_0(980)$ and $a_0(980)$ mesons, because
the definition of the nonet does not depend on the widths.
Therefore, these anomalies cannot serve as an argument for
introducing the exotic multiplet but should be explained on the
basis of the $q\bar{q}$ structure.

The flavor symmetry predicts that all particles belonging to the S
nonet form the straight flavor stitch line in the $(m,\Gamma)$
plane. The data show that the slope of the stitch line is
negative. The linearity may, however, be broken (for some
particles having masses smaller than $\simeq1.5\mathrm{GeV}$), if
the usual flavor-conditioned decay is distorted by some
"kinematical" mechanism. Such a kind of mechanism suppresses
decays of $f_0(980)$ and $a_0(980)$ mesons. We know some of its
properties: the suppression is strong; it does not depend on the
masses; it is independent of the isospin. But the present approach
does not identify its nature. The riddle of the scalar mesons
remains.

Obviously, with this "kinematical" mechanism suppressing the
$f_0(980)$ and $a_0(980)$ decays, investigation of the
$\delta^{I=0}_{J=0}$ and $\delta^{I=1}_{J=0}$ phases in the
resonance region does not yield information about properties of
the flavor symmetry.

Let us discuss now the decuplet. The decuplet is a real object -
 as real, as the nonet. It is a multiplet whose isoscalar octet
state is distributed among three isoscalar physical states. Its
properties, as well as properties of the nonet, are defined by ME.
Number of equations may be chosen in such a way that masses
satisfy the mass formula. With such a choice, the mass formula
plays the role of necessary condition of solvability of the ME.
The formula does not depend on the widths of the particles. The
solution of the ME, $|l_i|^2$ (i=1,2,3), serves for constructing
the mixing matrix of the decuplet. The matrix is based exclusively
on the solution and is real for real masses as well as for complex
ones. Its elements depend only on the masses. The particles of the
decuplet states form the straight flavor stitch line on the
$(m,\Gamma)$ plane, just as do the S nonet ones.

The decuplet includes the mesons $a_0(1450)$, $K_0(1950)$,
$f_0(1370)$ and $f_0(1500)$. The missing $f_0$ has a mass
somewhere in the region $2200\div2400$ MeV. The mass is strongly
correlated with the mass of $K_0(1950)$. Its vagueness reflects
inaccuracy of the $K(1950)$ mass and the error of the $a_0(1450)$
one. This meson is almost a pure $s\bar{s}$ state; therefore the
state G is almost completely distributed between the $f_0(1370)$
and $f_0(1500)$ mesons. The G content of each of these particles
is determined by relations between  $f_0(1370)$, $a_0(1450)$ and
$f_0(1500)$ masses. Thus the knowledge of these three masses is
sufficient for approximate evaluating the whole mixing matrix.

The fit of the flavor stitch line should be a necessary element of
the present investigation. One can expect good fit, as the masses
of the decuplet particles are large enough. The extra information
about the widths would be especially welcome for $f_0(1370)$ and
$f_0(1500)$ mesons which are expected to include a glueball.
However, the current data are too crude for that. More data and
better understanding of the decay processes are desirable \cite
{Barnes}.

In general, \it{consistency of the mixing parameters predicted
from the decuplet masses with the results of the analysis of the
isoscalar mesons production and decay would provide the requested
ultimate evidence of G}.\rm

The mass regions of the nonet and decuplet are overlaping.
However, such a situation should not be treated as an obstacle for
accepting the proposed distribution of the particles between the
multiplets. We may prefer to follow data rather than habitual
mixing the adjacent states. The $0^{++}$ nonet mesons well satisfy
the S mass formula. This suggests that mixing between the nonet
and decuplet states is negligible. A similar situation can be seen
for $1^{--}$ multiplets: the ground state nonet $(\rho, K^*,
\omega, \phi)$ is ideally mixed, while the higher lying states
$(\rho(1450), K^*(1410), \phi(1420))$ form the octet of exact
symmetry \cite{SR}. That could not happen if there were a
remarkable mixing between these multiplets. We can also notice
that for many $J^{PC}$ not only the ground state multiplet is
observed, but also the higher one; it would be impossible to
distinguish separate multiplets if the mixing were strong.

For the S nonet and decuplet, apart from the mass formulae, there
exist other necessary solvability conditions of the ME. They have
a form of the mass ordering rule. For the $0^{++}$ nonet we have
\begin{displaymath}
x_1<a<x_2<b,
\end{displaymath}
while for the $0^{++}$ decuplet it is
\begin{displaymath}
x_1<a<x_2<b<x_3.
\end{displaymath}
These rules are very useful in investigating the nonet and
decuplet of the scalar multiplets.

They throw also some light on the problem of $\sigma(600)$ meson.
According to these rules, the nonet of $0^{++}$ mesons cannot be
transmuted into a decuplet by joining the scalar meson with a mass
smaller than the one of the $f_0(980)$. Therefore, $\sigma(600)$
cannot be considered as a decuplet component and is a separate
state. It may be a genuine particle state (then it would be the
ground state unitary singlet), or it may be a state of a different
kind. A possibility that nature of the $\sigma(600)$ is different
than the nature of other scalar mesons has been discussed for some
time \cite{Pen}, \cite{WOchs}, \cite{Anisov}.

\section{Conclusion}
1. The most complete description of the meson multiplets (nonet
and decuplet) is given by the master equations (ME) which are
derived from the hypothesis about vanishing of the exotic
commutators. For the finite-width mesons they reveal the features
which were not known before. These features enable us to
understand the mass spectrum of the scalar mesons.

2. The $0^{++}$ mesons form the nonet ($a_0(980)$, $K_0(1430)$,
$f_0(980)$, $f_0(1710)$), the decuplet ($a_0(1450)$, $K_0(1950)$,
$f_0(1370)$, $f_0(1500)$, $f_0(2200\div2400)$) and a separate
state $\sigma(600)$.

3. There are no $q^2\bar{q}^2$ exotics. The nonet mesons satisfy
the Schwinger mass formula and are the usual $q\bar{q}$ states.
Anomalies of the $f_0(980)$ and $a_0(980)$ widths are caused by
some "kinematical" mechanism which suppresses their decay. The
energy dependence of the phases $\delta_{J=0}^{I=0}$ and
$\delta_{J=0}^{I=1}$ do not reflect properties of the flavor
interaction. The nature of the suppression mechanism remains
unknown.

4. The decuplet includes the glueball state. The mass formula and
mixing matrix of the decuplet isoscalar physical states follow
directly from the solution of the ME. The glueball is included in
the states of $f_0(1370)$ and $f_0(1500)$ mesons. Its contribution
to these states is completely determined by the masses of decuplet
particles. Agreement between quark-glueball structures of the
isoscalar physical states, determined in this way, and their
production and decay patterns would provide the ultimate
identification of the glueball.

5. The meson $\sigma(600)$ cannot be mixed with the nonet
(($a_0(980)$, $K_0(1430)$, $f_0(980)$, $f_0(1710)$) to form a
decuplet. This may support the conjecture about the peculiar
nature of this particle.

\section{Acknowledgments}
It is a pleasure for me to thank Profs S. B. Gerasimov, P.
Kosinski and V. A. Meshcheryakov for valuable discussions and
Profs P. Maslanka and J. Rembielinski for support during
completing this work.


\begin{thebibliography}{2}

\bibitem{Morgan} D. Morgan, {Phys. Lett.} {\bf 51B}, 71 (1974)
\bibitem{PDG} S. Eidelman, {Phys. Lett. B} {\bf 592}, 1 (2004)
\bibitem{G-I} S. Godfrey, N. Isgur, {Phys. Rev. D} \textbf{32}
189 (1985)
\bibitem{Jaffe} R. J. Jaffe, {Phys. Rev.} {\bf D15}, 267, 281 (1977)
\bibitem{Wein-Isg} J. Weinstein, N. Isgur, {Phys. Rev. Lett.} {\bf 48} 659
(1982); {Phys. Rev. D} {\bf 27} 588 (1983); {\bf 41} 2236 (1990)
\bibitem{Achas} N. N. Achasov, S. A. Devyanin, G. N. Shestakov {Z. Phys.
C - Particles and Fields} {\bf 22}, 53 (1984)
\bibitem{Tornq} N. A. Tornqvist, {Phys. Rev. Lett.} {\bf 49}, 624
(1982); {Z. Phys.  C - Particles and Fields} {\bf 68}, 647 (1995);
{hep-ph/9504372}
\bibitem{C-T} F. Close, N. Tornqvist, {J. Phys. G: Nucl. Part.
Phys} {\bf 28}, R249 (2002); {hep-ph/0204205}
\bibitem{SR} M. Majewski, {Eur. Phys. J. C} {\bf30}, 223
(2003); {hep-ph/0206285}
\bibitem{Ba} G. Bali et al., {Phys. Lett. B} \textbf{309}, 378 (1993),
{hep-lat/9304012}; {Phys. Rev. D} \textbf{62}, 054503 (2000)
\bibitem{Ger} S. S. Gershtein, A. K. Likhoded, Yu. D. Prokoshkin,
{Z. Phys. C} \textbf{24} 305 (1984)
\bibitem{Ani} V. V. Anisovich al., {Phys. Lett. B} \textbf{323}
233 (1994)
\bibitem{Am-Ar} C. Amsler et al., {Phys. Lett. B} \textbf{342} 433
(1995)
\bibitem{Am-C} C. Amsler, F. E. Close, {Phys. Lett. B}
\textbf{353} 385 (1995), {hep-ph/9505219}; {Phys. Rev. D}
\textbf{53} 295 (1996), {hep-ph/9507326}
\bibitem{C-K} F. E. Close, A. Kirk, {Phys. Lett. B} \textbf{483}
3445 (2000), {hep-ph/0004241}
\bibitem{CAms} C. Amsler, {Phys. Lett. B} \textbf{541} 22 (2002);
{hep-ph/0206104}
\bibitem{C-Z} F. E. Close, Q. Zhao, {Phys. Rev. D}  {\textbf71} 094022 (2005);
{hep-ph/0504043}
\bibitem{MT1} M. Majewski, W. Tybor, {Acta Physica Polonica B} {\bf 15}, 267 (1984)
\bibitem{MT2} M. Majewski, W. Tybor, {Acta Physica Polonica B}
{\bf 15}, 777 (1984); Erratum, {Acta Physica Polonica B} {\bf 15
$N^012$}, page 3 of the cover
\bibitem{KM} B. Kozlowicz, M. Majewski, {Acta Physica Polonica B}
{\bf 20}, 869 (1989)
\bibitem{In} M. Majewski, {Z. Phys. C} {\bf 39}, 121 (1988)
\bibitem{Freund} G.O.Freund, Y. Nambu, {Phys. Rev. Lett}
\textbf{34} 1645 (1975); N. Fuchs, {Phys. Rev. D} \textbf{14} 1912
(1976); D. Robson, {Nuclear Physics B} \textbf{130} 328 (1977); F.
E. Close, A. Kirk, {Euro Phys. J. C} \textbf{21} 531 (2001),
{hep-ph/0103173}
\bibitem{isn't} M. Majewski, W. Tybor, {Acta Physica Polonica B}
{\bf 17}, 333 (1986); Proc. of the IX Warsaw Symposium on
Elementary Particle Physics, Kazimierz, Poland, May 25 - 31, 1986
\bibitem{Aj-Wa-Ka} M. Majewski, The IIIrd Int. Conf. on Hadron
Spectroscopy "HADRON'89", Ajaccio, Corsica (France) Sept. 23-27
1989; "HADRON'91", College Park, 12-16 August 1991; Proc. of the
XV Int. Warsaw Meeting on Elementary Particle Physics, Kazimierz,
Poland, 25-29 May 1992
\bibitem{Dubna} M. Majewski, {Communication of the Joint Institute
of Nuclear Research}, {E2-91-546}, Dubna 1991
\bibitem{Prot} M. Majewski, {Thesis}, IHEP Protvino 1995, Russia

\bibitem{Gold} L. Burakovsky, T. Goldman {Phys. Rev. D}
\textbf{57} 2879 (1998), {hep-ph/9703271}; {Nucl. Phys. A}
\textbf{628} 87 (1998), {hep-ph/9709305}
\bibitem{Barnes} T. Barnes, {Invited talk at "Hadron '03"
Aschaffenburg, Germany 2003}, {hep-ph/0311102}
\bibitem{Pen} M. R. Pennington, {"Frascati 1999, Hadron Spectroscopy"
Talk given at Workshop on Hadron spectroscopy (WHS 99),
Roma,Italy}, {hep-ph/9905241}
\bibitem{WOchs} P. Minkowski, W. Ochs {Nucl. Phys. Proc. Supp.} \textbf{121} 123
(2003); {hep-ph/0209225}; W. Ochs, {Invited talk at "Hadron '03",
Aschaffenburg, Germany 2003}; {hep-ph/0311144}
\bibitem{Anisov} V. V. Anisovich, {Usp. Fiz. Nauk} {\textbf{47} 49
(2004)}, {hep-ph/0208123}

\end{thebibliography}
\end{document}